\documentstyle[aps, manuscript]{revtex}
\tightenlines
\draft
\begin{document}
\title{Inflaton field fluctuations from gauge-invariant metric fluctuations
during inflation}
\author{$^1$Mariano Anabitarte\footnote{
E-mail address: anabitar@mdp.edu.ar}
and $^{1,2}$Mauricio Bellini\footnote{
E-mail address: mbellini@mdp.edu.ar}}
\address{$^1$Departamento de F\'{\i}sica, Facultad de Ciencias
Exactas y Naturales
Universidad Nacional de Mar del Plata,
Funes 3350, (7600) Mar del Plata, Buenos Aires, Argentina.\\
and \\
$^2$ Consejo Nacional de Investigaciones Cient\'{\i}ficas y
T\'ecnicas (CONICET)}

\vskip .5cm
\maketitle
\begin{abstract}
The evolution of the inflaton field fluctuations
from gauge-invariant metric fluctuations is discussed.
In particular, the case of a symmetric $\phi_c$-exponential
potential is analyzed.
\end{abstract}
\vskip .2cm                             
\noindent
Pacs numbers: 98.80.Cq \\
\vskip 1cm

\section{Introduction}

An attractive proposal concerning the first moments of the observable
universe is that of chaotic inflation\cite{li}.
At some initial epoch, presumably
the Planck scale, the scalar field existing in nature are roughly homogeneous
and dominate the energy density. Their initial values are random,
subject to the constraint that energy density is at the Planck scale.
Among them is the inflaton field $\varphi$, which is distinguished
from the noninflaton fields by the fact that the potential is relatively
flat in its direction. This field would be responsible for the
inflationary expansion of the universe.
Inflationary model\cite{Guth}
solves several difficults which arise from the standard
cosmological model, such as the horizon, flatness, and monopole problems.
Furthermore, it provides a mechanism for the creation of primordial density
fluctuations needed to explain the structure formation in the universe.
Stochastic inflation\cite{yi,hm,habib,mijic,BCMS,riotto}
is a very interesting approach to inflation that
has played an important role in inflationary
cosmology in the last two decades. This approach gives the possibility
of making a description of the matter field fluctuations in the
infrared (IR) sector by means of the coarse-grained matter field\cite{Be},
that describes the inflationary universe on cosmological (super Hubble)
scales.
Since these perturbations are classical on super Hubble scales,
in this sector one can make a standard stochastic treatment for the
coarse-grained inflaton field. The IR sector is very important because
the spatial inhomogeneous would explain the present day observed
matter structure in the universe at cosmological scales. Matter fluctuations
are responsible for metric fluctuations in the universe around
the background Friedmann-Robertson-Walker (FRW) metric.
The theory of linearized gravitational perturbations in an expanding
universe is a very important subject of study in modern cosmology. It is used
to describe the growth of structure in the universe, to calculate the
predicted microwave background fluctuations, and
in many other considerations.
The growth of perturbations in an expanding universe is a consequence of
gravitational instability. A small overdensity will exert an extra
gravitational attractive force on the surrounding matter. Consequently,
the perturbation will increase and will in turn produce a larger attractive
force. In an expanding universe the increase in force is partially
counteracted by the expansion. This, in general, results in power-law growth
rather than exponential growth of the perturbations.
Mathematically, the problem of describing the growth of small perturbations
in the context of general relativity reduces to solving the Einstein
equations linearized on an expanding background\cite{1}.

In particular, the inflaton
field fluctuations are responsible for metric fluctuations
around the background FRW metric. When
metric fluctuations do not depend
on the gauge, the perturbed globally flat
isotropic and homogeneous universe is described by\cite{1}
\begin{equation}\label{m}
ds^2 = (1+2\psi) \  dt^2 - a^2(t) (1-2\Phi) \  dx^2,
\end{equation}
where $a$ is the scale factor of the universe and ($\psi$, $\Phi$) are
the gauge-invariant (GI) perturbations of the metric.
In the particular
case where the tensor $T_{\alpha\beta}$ is diagonal, one obtains:
$\Phi = \psi$\cite{1}. The field $\Phi$ is called relativistic potential
and describes the scalar GI metric fluctuations.
The coordinate system
(\ref{m}) is more convenient for the investigation of density perturbations
than the usual synchronous system.

We consider a semiclassical expansion for the
inflaton field $\varphi(\vec x,t) = \phi_c(t) + \phi(\vec x,t)$\cite{BCMS},
with expectation values
$\left<0|\varphi|0\right> = \phi_c(t)$ and $\left<0|\phi|0\right>=0$. Here,
$\left.|0\right>$ is the vacuum state.
Due to $\left<0|\Phi|0\right> =0$, the expectation
value of the metric (\ref{m}) gives the background
metric that describes a flat FRW spacetime: $\left<ds^2\right>
= dt^2-a^2 dx^2$.

The Einstein equations can be linearized in terms of $\phi$ and $\Phi$
and the resulting equations for matter and metric fluctuations are
\begin{eqnarray}
\frac{1}{a^2} &\nabla^2 & \Phi - 3 H \dot\Phi - \left(
\dot H + 3 H^2\right) \Phi = \frac{4\pi}{M^2_p} \left(
\dot\phi_c \dot\phi + V' \phi\right), \label{a}\\
\frac{1}{a}& \frac{d}{dt}& \left( a \Phi \right) = 
\frac{4\pi}{M^2_p} \left(\dot\phi_c \phi\right) , \label{3} \\
\ddot\phi& +& 3 H \dot\phi -
\frac{1}{a^2} \nabla^2 \phi + V''(\phi_c) \phi 
+ 2 V'(\phi_c) \Phi- 4 \dot\phi_c \dot\Phi =0, \label{A}
\end{eqnarray}
where $a$ is the scale factor of the universe and
prime and overdots denote respectively the derivatives with respect to
$\phi_c$ and time. Furthermore,
and $H=\dot a/a$ give us the Hubble parameter.

In this paper we are aimed to study the evolution
of the inflaton field fluctuations $\phi$ 
from GI metric fluctuations. To make it, firstly we must to solve
the equations (\ref{a}) and (\ref{A}) for $\Phi$, to be able
the solution for $\phi$ in eq. (\ref{3}).

\section{Inflaton fluctuations from GI metric fluctuations}

The dynamics of $\phi_c$ on the background FRW metric
is given by the equations
\begin{equation}\label{dyn}
\ddot\phi_c + 3 H\dot\phi_c + V'(\phi_c)=0,
\qquad \dot\phi_c = -\frac{M^2_p}{ 4\pi} H'.
\end{equation}
Furthermore, the scalar potential can be written in terms of
the Hubble parameter
\begin{equation}
V(\phi_c) = \frac{3 M^2_p}{8\pi} \left[H^2 - \frac{M^2_p}{12\pi} \left(
H'\right)^2\right].
\end{equation}
If we Replace eq. (\ref{3}) in (\ref{a}), we obtain the Klein-Gordon
like equation for the GI metric fluctuations $\Phi$
\begin{equation}\label{1}
\ddot\Phi + \left(H
- 2 \frac{\ddot\phi_c}{\dot\phi_c} \right)
\dot \Phi - \frac{1}{a^2} \nabla^2 \Phi +2\left(
\dot H - H \frac{\ddot\phi_c}{\dot\phi_c}\right) \Phi =0.
\end{equation}
The equation (\ref{1}) can be simplified by introducing
the redefined field $Q = e^{1/2\int\left[H - 2\ddot\phi_c/\dot\phi_c\right]dt}
\Phi$. If we make it, we obtain
\begin{eqnarray}
\ddot Q &-& \frac{1}{a^2} \nabla^2 Q -
\left[
\frac{1}{4} \left( H
- 2 \frac{\ddot\phi_c}{\dot\phi_c}\right)^2
+\frac{1}{2}\left[
\dot H- 2\frac{d}{dt}\left(\frac{\ddot\phi_c}{ \dot\phi_c}\right)\right]
-2 \left(\dot H -H \frac{\ddot\phi_c}{\dot\phi_c}\right)
\right] Q
= 0\label{h}.
\end{eqnarray}
This field can be expanded in terms of the modes
$Q_k=e^{i\vec k.\vec x} \xi_k(t)$
\begin{equation}
Q(\vec x,t) = \frac{1}{(2\pi)^{3/2}} {\Large \int} d^3k \left[
\alpha_k Q_k(\vec x,t) + \alpha^{\dagger}_k Q^*_k(\vec x,t)\right],
\end{equation}
where $\alpha_k$ and $\alpha^{\dagger}_k$ are the annihilation
and creation operators that complies with the commutation relations
\begin{eqnarray}
\left[\alpha_k,\alpha^{\dagger}_{k'} \right]&=& \delta^{(3)}(k-k'), \label{c1}\\
\left[\alpha_k,\alpha_{k'} \right]&=&\left[
\alpha^{\dagger}_k,\alpha^{\dagger}_{k'} \right]=0.\label{c2}
\end{eqnarray}
The equation for the modes $Q_k$ is
\begin{equation}\label{xxi}
\ddot{Q}_k + \omega^2_k(t) \  Q_k =0,
\end{equation}
where $\omega^2_k = a^{-2}\left(k^2 - k^2_0\right)$ is the squared
time dependent frequency for each $k$-mode
and $ k_0$ separates the infrared ($k\ll k_0(t)$)
and ultraviolet ($k \gg k_0(t)$) sectors
\begin{equation}
\frac{k^2_0}{a^2} =
\frac{1}{4} \left( H
- 2 \frac{\ddot\phi_c}{\dot\phi_c}\right)^2
+\frac{1}{2}\left[
\dot H- 2\frac{d}{dt}\left(\frac{\ddot\phi_c}{ \dot\phi_c}\right)\right]
-2 \left(\dot H -H \frac{\ddot\phi_c}{\dot\phi_c}\right)
\end{equation}
Furthermore, the field $Q$ obeys the 
the following commutation law:
$\left[Q(\vec x,t)\dot Q(\vec x',t)\right]
= i \delta^{(3)}\left(\vec x-\vec x'\right)$,
so that the modes $\xi_k$ are renormalized by the expression
\begin{equation}\label{ren}
\dot \xi^*_k \xi_k - \dot \xi_k \xi^*_k = i.
\end{equation}

The inflaton field oscillates around 
the minimum of the potential
at the end of inflation. Due to this fact the solutions of the
eq. (\ref{xxi}) when $\dot\phi_c =0$
and $\ddot\phi_c=0$ are very important.
When $\dot\phi_c=0$ we obtain that $Q_k=0$, but the
solutions for $\Phi_k$ are given by
\begin{equation}
\Phi_k = a^{-1} \Phi^{0}_k.
\end{equation}
where $\Phi^{0}_k$ is the initial amplitude for $\Phi_k$, for each
wavenumber $k$. This means that the amplitude of each mode $\Phi_k$
decreases with the expansion of the universe.
On the other hand, when $\ddot\phi_c=0$, the field
is at the minimum of the potential.
At this moment the equation (\ref{xxi}) takes the form
\begin{equation}
\ddot Q_k
+  \left[ \frac{k^2}{a^2} - \left(\frac{H^2}{4} - \frac{3}{2} \dot H
\right)\right] Q_k =0,
\end{equation}
where $\Phi_k = a^{-1/2} Q_k$.

Now we can write the equation (\ref{3}) in terms of the field $Q$.
Once we know the modes $Q_k$, the modes $\phi_k$ for the
gauge-invariant inflaton fluctuations will be determined by
\begin{equation}\label{if}
\phi_k = \frac{M^2_p}{4\pi a \dot\phi_c} e^{-\frac{1}{2} \int \left[
H-2\frac{\ddot\phi_c}{\dot\phi_c}\right] dt} \left\{
Q_k \left[ \frac{H}{2} + 2\frac{\ddot\phi_c}{\dot\phi_c}\right]
+ \dot Q_k \right\}.
\end{equation}
Hence, if we assume that slow-roll conditions\cite{copeland} are fulfilled
(it should be before the reheating period),
the fluctuations for energy density will be
\begin{equation}
\frac{\delta\rho}{\rho} \simeq \frac{V'}{V} \left< \phi^2\right>_{GI},
\end{equation}
where the squared fluctuations of $\phi$ are
\begin{equation}
\left<\phi^2\right>_{GI}= \frac{1}{2\pi^2} {\Large\int} dk \  k^2
\phi_k(t) \phi^*_k(t).
\end{equation}
Here, the modes $\phi_k$ are given by the eq. (\ref{if}).

\section{power-law inflation
for a symmetric $\phi_c$-potential}

To ilustrate the formalism we can examinate a scalar potential
given by $V(\phi_c) = V_0 \  e^{2\alpha |\phi_c|}$, where
$\alpha^2={4\pi \over M^2_p p}$ gives the relationship
between $\alpha $ and the power of the expansion $p$
for a scale factor that increases as $a \sim
t^p$.
The Hubble parameter is given by $H(t)=p/t$, or, in term of $\phi_c$
\begin{equation}
H_c =\frac{\pi}{ M_p}
\left(\frac{32 V_0}{12\pi - \alpha^2 M^2_p}
\right)^{1/2} \  e^{\alpha |\phi_c|},
\end{equation}
where $V_0 = {3 M^2_p \over 8\pi} H^2_e \left[{12\pi - M^2_p \alpha^2
\over 12\pi}\right]$ and $H_e=p/t_e$ is the value of the Hubble
parameter at the end of inflation.
Furthermore, the evolution for $|\phi_c(t)|$ is 
\begin{equation}
|\phi_c(t)| = |\phi_0 | - \frac{1}{\alpha} {\rm ln}\left(\frac{t}{t_0}\right),
\end{equation}
where $t \geq t_0$. Since $\dot\phi_c = -{\rm sgn}(\phi_c){1\over \alpha t}$
and $\ddot\phi_c = {\rm sgn}(\phi_c){1\over \alpha t^2}$
[we assume ${\rm sgn}(\phi_c)= \pm 1$ for $\phi_c$ positive and
negative, respectively], the
evolution for $\Phi$ will be described by the equation
\begin{equation}
\ddot\Phi + \frac{(p+2)}{t} \dot\Phi - \frac{1}{a^2} \nabla^2 \Phi =0.
\end{equation}
Now we can make the transformation $Q = \Phi e^{\int (p+2)t^{-1} dt}$
so that the differential equation for $Q$ yields
\begin{equation}
\ddot Q - \frac{1}{a^2} \nabla^2 Q - \left[\frac{p}{2}\left(\frac{p}{2}+1
\right)t^{-2} \right]Q=0.
\end{equation}
The general solution for the modes $Q_k(t)$ is
\begin{equation}
Q_k(t) = C_1 \sqrt{\frac{t}{t_0}} {\cal H}^{(1)}_{\nu}[x(t)]
+ C_2 \sqrt{\frac{t}{t_0}} {\cal H}^{(2)}_{\nu}[x(t)],
\end{equation}
where ($C_1$,$C_2$) are constants, (${\cal H}^{(1)}_{\nu}[x]$,
$H^{(2)}_{\nu}[x]$) are the Hankel functions of (first, second) kind
with $x(t) = {t^p_o k \over a_0 (p-1) t^{p-1}}$ and $\nu = {p+1 \over
2(p-1)}$. Using the renormalization condition (\ref{ren}),
we obtain the Bunch-Davis vacuum\cite{BD} solution
($C_1=0$,$C_2 = \sqrt{\frac{\pi}{2(p-1)}}$) 
\begin{equation}\label{H}
Q_k(t) = \sqrt{\frac{\pi}{2}} \sqrt{\frac{t}{t_0(p-1)}}
{\cal H}^{(2)}_{\nu}[x(t)],
\end{equation}
In the UV sector the function ${\cal H}^{(2)}_{\nu}[x]$
adopts the
asymptotic expression (i.e., for $x \gg 1$)
\begin{equation}\label{H1}
{\cal H}^{(2)}_{\nu}[x] \simeq \sqrt{\frac{2}{\pi x}}\left[
{\rm cos}\left(x-\nu\pi/2 - \pi/4\right)-i \  {\rm sin}
\left(x-\nu\pi/2 - \pi/4\right)\right],
\end{equation}
whilst on the IR sector (i.e., for $x \ll 1$)
it tends asymptotically to
\begin{equation}\label{H2}
{\cal H}^{(2)}_{\nu}[x] \simeq \frac{1}{\Gamma(\nu+1)} \left(
\frac{x}{2}\right)^{\nu} - \frac{i}{\pi} \Gamma(\nu) \left(
\frac{x}{2}\right)^{-\nu}.
\end{equation}
The $\Phi$-squared field fluctuations on the IR sector are
$\left(\left<\Phi^2\right>\right)_{IR} =
{1 \over 2\pi^2} {\Large\int}^{\epsilon k_0(t)}_{0} dk k^2
\left|\Phi_k\right|^2$, and becomes
\begin{eqnarray}
\left(\left<\Phi^2\right>\right)_{IR} & \simeq &
\frac{1}{4}\left\{\frac{\left[\frac{t^p_0}{2 a_0(p-1)}\right]^{2\nu}
\left[\frac{\epsilon a_0 \sqrt{\frac{p}{2}\left(\frac{p}{2}+1\right)}}{t^p_0}
\right]^{\frac{2(2p-1)}{p-1}}}{
\pi t_0 \Gamma^2\left(\frac{3p-1}{2(p-1)}\right)(3p-1)}\right. \nonumber \\
& + & \left.
\frac{\Gamma^2\left(\nu\right) \left[\frac{t^p_0}{
2 a_0(p-1)}\right]^{-2\nu} \left[\frac{\epsilon a_0 \sqrt{
\frac{p}{2}\left(\frac{p}{2}+1\right)}}{t^p_0}\right]^{\frac{2(p-2)}{p-1}}}{
\pi^3 t_0 (p-3)} \right\}  t^{3-2\nu},\label{AA}
\end{eqnarray}
where $\epsilon =k^{(IR)}_{max} /k_p \ll 1$ is a dimensionless constant,
$k^{(IR)}_{max} = k_0(t_*)$ at the moment $t_*$ when the horizon entry and
$k_p$ is the Planckian wavenumber (i.e., the scale we choose as a cut-off
of all the spectrum). The power spectrum on the IR sector is
$\left.{\cal P}_{\Phi}\right|_{IR} \sim k^{3-2\nu}$.
Note that  $\left(\left<\Phi^2\right>\right)_{IR}$
increases for $p >2$, so that to the IR squared $\Phi$-fluctuations
remain almost constant on cosmological scales
we need $p \simeq 2$. We find that
a power close to $p =2$ give us a scale invariant power
spectrum (i.e., with
$\nu \simeq 3/2$ for $\left(\left<\Phi^2\right>\right)_{IR}$.
Furthermore, density fluctuations for matter energy density are given
by $\delta\rho/\rho
= -2\Phi$, so that $\left<\delta\rho^2\right>^{1/2}/\left<
\rho\right> \sim \left<\Phi^2\right>^{1/2}$.

On the other hand, in the UV sector these fluctuations are given by
\begin{equation} \label{BB}
\left(\left<\Phi^2\right>\right)_{UV} \simeq
\frac{a_0}{4 t^{p+1}_0 \pi^2}
\left\{\frac{k^2_p}{t^2} -
\frac{a^2_0}{t^{2p}}
\left[\frac{p}{2} \left(\frac{p}{2}+1\right)\right]\right\}
t^{3-2\nu}.
\end{equation}
The power spectrum in this sector go as $\left.{\cal P}_{\Phi}\right|_{UV}
\sim k^4$.
We observe from eq. (\ref{BB})
that $\left(\left<\Phi^2\right>\right)_{UV}$
increases during inflation for $p > 3$. 
From the results (\ref{AA}) and (\ref{BB}) we obtain that
$1 < p \le 2$, because a power-law $p>2$
could give a very inhomogeneous universe on cosmological scales.
Since $\left(\left<\Phi^2\right>\right)_{UV} \ge 0$,
we obtain the condition
\begin{equation}
k^2_p -
\frac{a^2_0}{t^{2(p-1)}} \left[\frac{p}{2} \left(\frac{p}{2}+1\right)\right]
\ge 0.
\end{equation}
If $a_0=H^{-1}_0$ ($H_0$ is the initial value of the Hubble parameter),
inflation should ends at $t=t_e$, where
\begin{equation}\label{uaa}
t_e \simeq \left[\frac{k_p H_0}{\sqrt{\frac{p}{2}\left(\frac{p}{2}+1\right)}}
\right]^{\frac{1}{p-1}}.
\end{equation}
For example, for $k_p H_0 = 10^{11} \  M_p$ and $p=2$, we
obtain $t_e \simeq 5.8 \  10^{10} \  M^{-1}_p$.

Furthermore, from eq. (\ref{if}) we obtain the solutions for the modes
$\phi_k(t)$
\begin{equation}
\phi_k(t) = {\rm sgn}(\phi_c) M_p  \sqrt{\frac{1}{8 t_0 p (p-1)}}
\left[ t^{-(p+1)/2} {\cal H}^{(2)}_{\nu}[x(t)] -
k \frac{t^p_0}{a_0} t^{-(3p-1)/2} {\cal H}^{(2)}_{\nu +1}[x(t)]\right],
\end{equation}
where ${\cal H}^{(2)}_{\nu}[x(t)]$ takes the asymptotic expressions
(\ref{H1}) and (\ref{H2}) for
$x \gg 1$ and $x \ll 1$, respectively.
The $k$-modes for the inflaton field fluctuations on
the IR and UV sectors are given respectively by
\begin{eqnarray}
\left(\phi_k(t)\right)_{IR} & \simeq &
{\rm sgn}(\phi_c) M_p \sqrt{\frac{1}{8 t_0 p(p-1)}} \left\{
\frac{1}{\Gamma(\nu +1)} \left(\frac{t^p_0}{2 a_0 (p-1)} \right)^{\nu}
t^{-(p+1)} k^{\nu} \right. \nonumber \\
& -  & \frac{1}{\Gamma(\nu +2)} \left(\frac{t^p_0}{
2 a_0 (p-1)}\right)^{\nu+1} t^{1-3p} k^{\nu+2} \frac{t^p_0}{a_0} \nonumber \\
& + & \left.
\frac{i}{\pi} \left[ \Gamma(\nu) \left(\frac{2 a_0(p-1)}{t^p_0}\right)^{\nu}
- \Gamma(\nu+1) \left(\frac{2 a_0 (p-1)}{t^p_0}\right)^{\nu +1} \right]
k^{-\nu} \right\}, \\
\left(\phi_k(t)\right)_{UV} & \simeq &
{\rm sgn}(\phi_c) \frac{M_p}{2} \sqrt{\frac{a_0}{p t^{p+1}_0 \pi}}
\left[ t^{-1}
e^{-i\left[\frac{t^p_0 k}{a_0 (p-1) t^{p-1}} -\frac{\pi}{2(p+1)}
\right]} k^{-1/2} - \left(\frac{t^p_0}{a_0}\right) t^{-p}
e^{-i \left[\frac{t^p_0 k}{a_0 9p-1) t^{p-1}}-
\frac{\pi(2p-1)}{2(p-1)} \right]} k^{1/2}\right],
\end{eqnarray}
so that the squared $\phi$-fluctuations on both sectors are
\begin{eqnarray}
\left(\left< \phi^2 \right>\right)_{IR} & \simeq &
A \  t^{2(p-2)}, \\
\left(\left< \phi^2 \right>\right)_{UV} & \simeq &
B_1 \  t^{-2} + B_2 \  t^{-2p} + B_3 \  t^{-(p+1)} - B_4 \  t^{2(p-2)},
\end{eqnarray}
where the constants $A$, $B_1$, $B_2$, $B_3$ and $B_4$ are
\begin{eqnarray}
&& A= \frac{M^2_p}{8\pi^4 t_0(p-1)^2 p A^2_2 A^2_1 a^2_0} \left\{
\pi^2 A^{\frac{2p-1}{p-1}}_3 A^2_1 a^2_0 \left(\frac{t_0}{a_0(p-1)}
\right)^{\frac{p+1}{p-1}} \left[16^{\frac{p}{1-p}} p -2^{\frac{1-5p}{p-1}}
\right] \right.\nonumber \\
& + & a^2_0 A^{\frac{p-2}{p-1}}_3 A^2_2 A^2_1 \Gamma\left(\frac{p+1}{2(p-1)}
\right) \left[\frac{a_0}{t^p_0} (p-1)\right]^{\frac{p+1}{p-1}}
\left[ 2^{\frac{(5-p)}{p-1}} p - 16^{\frac{1}{p-1}}\right] \nonumber \\
& + & A^{\frac{p-2}{p-1}}_3 A^3_2 A^2_1 a^2_0 \Gamma\left(\frac{p+1}{2(p-1)}
\right) \left[ \frac{a^2_0}{t^{2p}_0}(p^2-p+1)\right]^{\frac{p}{p-1}}
\left[416^{\frac{1}{p-1}} - 2^{\frac{p+3}{p-1}} p \right] \nonumber \\
&+ & a^2_0 A^2_1 A^4_2 a^{\frac{p-2}{p-1}}_3 \left[\frac{a_0}{t^p_0} (p-1)
\right]^{\frac{3p-1}{p-1}} \left[2^{\frac{p+3}{p-1}} p - 4^{\frac{p+1}{p-1}}\right]
\nonumber \\
& + & \pi^2 \left(\frac{t^p_0}{a_0 (p-1)}\right)^{\frac{3p-1}{p-1}} t^{2p}_0
A^{\frac{4p-3}{p-1}}_3 A^2_2 \left[ 42^{\frac{(7-11p)}{p-1}} p -
32^{\frac{(7-11p)}{p-1}} \right] \nonumber \\
& + & \left.\pi^2 \frac{a_0}{t^p_0} 16^{\frac{(1-2p)}{p-1}}
\left(\frac{t_0}{a_0 (p-1)}\right)^{\frac{2p}{p-1}} f^{\frac{3p-2}{p-1}}_3
A_2 A_1 \left[2-3p\right]\right\}, \\
&& B_1 = \frac{M^2_p}{16 \pi^3 t^{p+1}_0 p} k^2_p, \\
&& B_2 = \frac{M^2_p t^{p-1}_0 k^4_p}{32 p \pi^3 a^2_0}, \\
&& B_3 = \frac{M^2_p}{12 \pi^3 t_0 a_0 p} \left(k^3_p
\frac{t^{p}_0}{a_0}\right) \\
&& B_4 = \frac{a^2_0}{t^{2p}} \left[\frac{1}{2} \left(\frac{p^2}{4}+\frac{p}{2}
\right) + \frac{1}{4} \left(\frac{p^2}{4}+ \frac{p}{2}\right)^2 +
\frac{2}{3} \left(\frac{p^2}{4} + \frac{p}{2}\right)^{3/2}\right],
\end{eqnarray}
where $A_1 = \Gamma\left(\frac{5p-3}{2(p-1)}\right)$,
$A_2 = \Gamma\left(\frac{3p-1}{2(p-1)}\right)$,
$A_3 = \frac{a^2_0}{t^{2p}_0}p(p+1)$ and
$k_p$ is the wave number at the Planckian scale.
Note that
for a scale invariant $\left(\left<\Phi^2\right>\right)_{IR}$ -
power spectrum with $p=2$ (i.e., for $\nu=3/2$),
the squared inflaton fluctuations on the infrared sector
$\left(\left<\phi^2\right>\right)_{IR}$
and the late times squared fluctuations
on the ultraviolet sector $\left(\left<\phi^2\right>\right)_{UV}$
are constant.

\section{Final Comments}

In this paper we have studied
the evolution of the inflaton field
fluctuations in a symmetric $\phi_c$-exponential
potential, from
GI metric fluctuations previously renormalized
by eq. (\ref{ren}).
Metric fluctuations
are here considered in the framework of the linear perturbative corrections.
The scalar metric perturbations are spin-zero projections of the graviton,
which only exists in nonvacuum cosmologies. The issue of gauge invariance
becomes critical when we attempt to analyze
how the scalar metric perturbations
produced in the early universe influence a background globally flat isotropic
and homogeneous universe. This allows us to formulate the problem of the
amplitude for the scalar metric perturbations on the evolution of the
background FRW universe in a coordinate-independent
manner at every moment in time. 
Note that we have not considered back-reaction effects
which are related to a second-order metric tensor fluctuations. 
In the power-law expanding universe here studied the GI metric
fluctuations are well described by the field $\Phi$
and predicts a scale invariant power
spectrum on the IR sector for $p = 2$\cite{AB}.
The interesting of the result here
obtained is that for $p=2$ the inflaton field
fluctuations result to be scale invariant on the IR
sector, but also on the UV sector if we consider
a cut-off $k_p$ on the Planckian scale.
Hence, the problem of the UV divergence for the inflaton
field fluctuations would be avoided, but also the problem of
the temporal increasing of such that fluctuations on small
scales, because  $\left(\left<\phi^2\right>\right)_{UV}$ becomes
squeezed at the end of inflation.

\vskip .2cm
\centerline{\bf{Acknowledgements}}
\vskip .2cm
MB acknowledges CONICET, AGENCIA 
and Universidad Nacional de Mar del Plata
for financial support.\\

\end{document}